# Band Edge Effects and Normal Mode Propagation in Waveguide Magnetophotonic Crystals

M. Levy, R. Li*, A. A. Jalali and X. Huang

Physics Department, Michigan Technological University, *1400 Townsend Drive, Houghton, MI 49931, U.S.A.*

*Materials Science Department, Michigan Technological University, *1400 Townsend Drive, Houghton, MI 49931, U.S.A.*

A study of band edge phenomena in ridge waveguide magnetophotonic crystals is presented. Normal mode analysis is used to examine the near-band edge and Bragg center-wavelength group velocities and their relation to the polarization. Large polarization departures from the input are observed in Bragg gratings patterned by focused-ion-beam milling on liquid-phase-epitaxial (LPE) $(Bi,Lu)_{(2.8 \pm 0.1)}Fe_{(4.6 \pm 0.2)}O_{(12.4 \pm 0.3)}$ films on (111) gadolinium gallium garnet (GGG) substrates. Transfer matrix analysis of the spectral response in the photonic crystals allows the determination of the phase of the elliptically polarized normal modes from measured transmittances and polarization rotation. Unlike non-birefringent Faraday rotators a non-linear-relation exists between polarization rotation and the normal mode phase difference. Large phase differences between right- and left-elliptically polarized normal modes and optical slow-down are found near the band edges and at the resonant cavity center-wavelength. Standing-wave photon trapping and optical slow-down are analyzed and compared for these systems.

**Key words:** magnetophotonic crystals, Faraday Effect, optical waveguides, magneto-optic films

## 1. Introduction

Band edge effects and resonant phenomena in magnetophotonic crystals are of interest because of their influence on polarization and speed of light control, of significant import in signal processing. Such structures may lead to magnetically tunable optical delay lines and optical buffers for signal routing or data storage.

The study of photonic crystal effects in magnetic materials has developed rapidly in the last few years and interest in this field continues to grow [1)-15)]. The present work addresses the development of magnetophotonic crystals in optical waveguide geometries. On-chip waveguide devices are appealing for a number of reasons besides the natural allure of monolithic integration. Of particular note is the possibility of attaining high index contrast in the photonic crystal through patterning of air gaps in the plane of the film and the enhancement of the coupling strength through high-order mode backscattering.[6)-8),16)]

## 2. Polarization in magnetophotonic waveguides and band edge effects

In this article we present a study of band edge phenomena in ridge waveguide magnetophotonic crystals as they relate to the manipulation of optical phases and speeds. Normal mode analysis is used to understand the effect of near-band edge group velocity on the polarization. Large polarization departures from the input are observed and their relation to normal mode speed modifications is addressed. Bragg gratings patterned by focused-ion-beam milling on 2.86-μm-thick liquid-phase-epitaxial (LPE) $(Bi,Lu)_{(2.8 \pm 0.1)}Fe_{(4.6 \pm 0.2)}O_{(12.4 \pm 0.3)}$ films on (111) gadolinium gallium garnet (GGG) substrates constitute the photonic crystals.

30kV gallium ions are used to pattern the photonic bandgap structure with period 338nm in length. Linear-birefringence in the slab is measured at 0.0006, 0.0046, and 0.0106 for the first three waveguide modes, from fundamental to second order, respectively.

200-μm-long gratings are formed on waveguide ridges to a specification of 600nm-deep grooves. 6μm-wide 800nm-high ridges fabricated by standard photolithography and dry-etching form the optical channels. Overall ridge waveguide lengths are 1.18 mm. Figure 1 shows a scanning-electron micrograph of the patterned photonic crystal ridge and a seven-quarter wave phase-shift resonant cavity step. The gratings are placed 20 μm away from the input facets to minimize polarization and birefringence effects from the waveguide segment before the light couples into the photonic crystal.

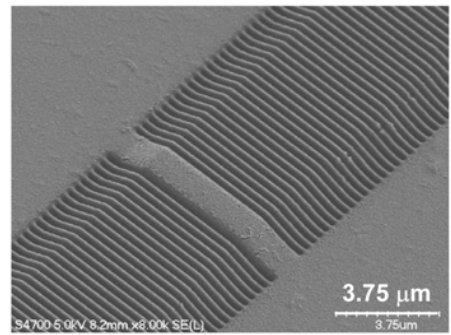

**Fig. 1** Scanning electron micrograph of magnetophotonic crystal on ridge waveguide channel. The image shows a portion of the grating containing the seven-quarter wave phase-shift step at the center of the structure.

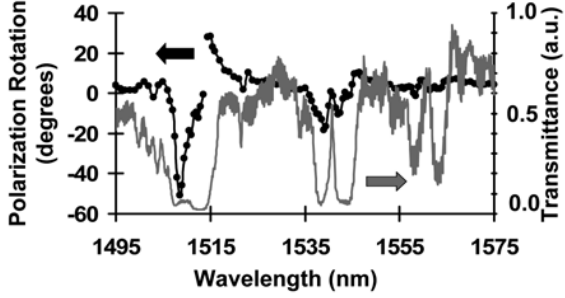

**Fig. 2** Transmittance and polarization rotation for one-dimensional magnetophotonic crystal on ridge waveguide with a single phase-shift step. Different stop bands correspond to backscatter coupling to fundamental through second-order waveguide modes.

Linearly polarized light from a tunable 1480-1580nm wavelength laser source is coupled into the waveguide by end-fire fiber coupling through a lensed fiber. Detection takes place by imaging the output signal into a photodetector through a 10× microscope objective. A saturating 50 Oe magnetic field is applied collinear to the waveguide axis and measurements are taken for parallel and anti-parallel directions of the field to the propagation direction.[12),13)] The polarization direction is defined as the direction of the semi-major axis of the ellipse, and the polarization rotation angle as half the difference between the output polarization directions corresponding to both magnetic fields. Insertion losses are estimated at 6dB.

Figure 2 displays the transmitted intensity and the direction of the semi-major axis of the output polarization ellipse relative to the input linear polarization direction for transverse electric (TE) input. Three stop bands are visible in these samples. Of particular interest here are the large departures from input polarization near the band edges in the short wavelength side of the spectral range. Figure 3 shows the transmittance and polarization rotation for first-order mode backscattering in a photonic crystal without phase shift step (Bragg grating) fabricated on the same garnet film under the same conditions described above. Also re-plotted for comparison (Fig. 3(a)) is the corresponding polarization rotation for first-order mode backscattering from Fig. 2. The large departures in polarization rotation near the band edge in Fig. 3(b) show that the polarization changes exist even in the absence of a resonant cavity. Thus group velocity variations and not solely photon trapping in a cavity (in the sense of standing waves) are responsible for the observed excursions in polarization. Differential variations in group velocity between the right- and left-elliptically polarized components of the propagating beam as a result of the resonant splitting between their respective bands and the shape of the band structure near the band edge can lead to additional output phase differences between them and increased polarization rotations. The spectral location of the polarization rotation peaks in Fig. 3(b) may be affected by the spread in the band edges.

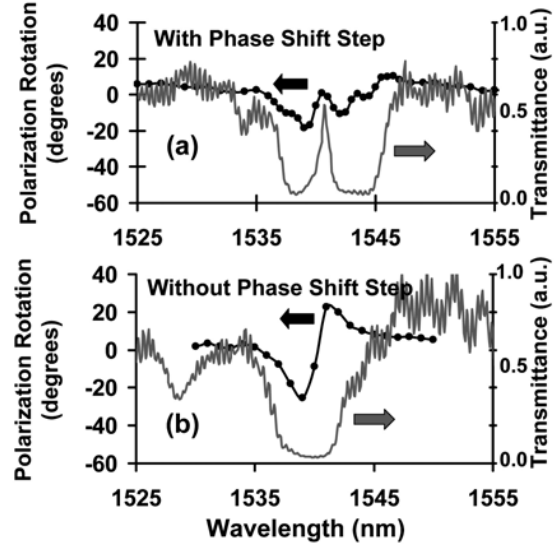

**Fig. 3** Polarization rotation for backscatter-coupling to the first-order waveguide mode for one-dimensional magnetophotonic crystals with (a) and without (b) phase-shift steps. More than 90% of the forward propagating power is in the fundamental mode. The polarization rotation enhancement in case (b) shows that there is a change in group velocity as a result of the photonic bandgap structure.

The stop band multiplicity in these photonic crystals evident in Fig. 2 is the result of inter-modal coupling between forward fundamental mode propagation and backscattering to high-order waveguide modes by the Bragg grating. Each of the stop bands visible in Fig. 2 results from Bragg scattering under the condition $\lambda = \Lambda(n_f + n_b)$. Here $\lambda$ is the optical wavelength in vacuum, $\Lambda$ the grating period, and $n_f$ and $n_b$ are the modal effective indices of the forward and backward propagating beams, respectively. We stress that the effect under discussion is not high-order Bragg scattering which occurs well outside the optical wavelength range under study but inter-modal coupling between different waveguide modes. Of especial interest is the strong coupling and large rotations that occur for fundamental forward to high-order backscattered modes in Fig. 2 and Fig. 3. These effects stem from an enhancement in the grating coupling strength for high-order mode backscattering.[16)]

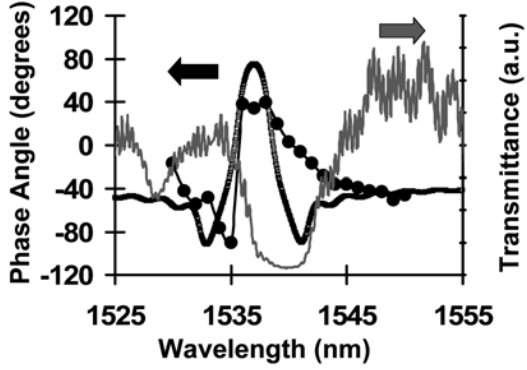

**Fig. 4** Phase-angle difference $\Delta\zeta$ between normal modes for an optical channel without phase shift step. The black circles are computed from the polarization-rotation and transmittance data. A curve computed via transfer-matrix simulation is also plotted for comparison. The transmittance is shown for reference.

### 3. Normal mode analysis

#### 3.1 Measured polarization rotation, phase retardation and normal modes

The large changes in polarization direction in the stop bands noted above can be understood in terms of accrued phase differences between normal modes in the optical guide. This, of course, is also the origin of the Faraday rotation in bulk magneto-optic systems, where the normal modes of propagation are circularly polarized waves with opposite sense of rotation traveling at different speeds. For magnetophotonic optical waveguide channels the normal modes become elliptically-polarized waves having opposite sense of rotation.[15]

The ratio between semi-major and semi-minor components in the magnetophotonic crystal normal modes is:[15]

$$\left(\frac{E_x^{(n)}}{E_y^{(n)}}\right)_\pm = i\,\frac{\alpha}{\frac{\beta}{2} \pm \sqrt{\alpha^2 + \frac{\beta^2}{4}}} \qquad (1)$$

In the above expression $\alpha$ is the Faraday rotation per unit length for the medium in question and $\beta$ is the linear birefringence retardation per unit length $2\pi(n_y - n_x)/\lambda$. This linear birefringence depends on the effective indices of the TE and TM modes in the waveguide. The x- and y- coordinates indicate the normal mode axes of linear birefringence retardation, $i=\sqrt{-1}$, the plus and minus signs denote the sense of rotation, and the superscript (n) stands for normal mode. The model assumes a lossless medium.

The direction $\theta$ of the semi-major axis of the output polarization ellipse is given by

$$\tan 2\theta = \frac{2x_0 y_0}{x_0^2 - y_0^2}\cos\phi \qquad (2)$$

Here $\phi$ is the phase difference between the y and x-components of the polarization, and $x_0, y_0$ are the magnitudes of the electric field vectors in the x- and y-directions. The latter can be measured with the help of a polarization analyzer at the output end. Thus $x_0, y_0,$ and $\theta$ are directly measurable quantities. The phase difference between the outgoing elliptical normal modes $\Delta\zeta = \zeta^+ - \zeta^-$ is related to the polarization rotation through the following expression:[15]

$$Exp(i\Delta\zeta) = \frac{|t^-|}{|t^+|}\frac{\left[i + z_0\left|\frac{E_y^{(n)}}{E_x^{(n)}}\right|_+\right]}{\left[i - z_0\left|\frac{E_y^{(n)}}{E_x^{(n)}}\right|_+\right]}, \qquad (3)$$

where $z_0 = [\,y_0/x_0\,]exp(i\phi)$. In the above expression, $t^\pm$ are the complex transmission coefficients for the structure:

$$t^\pm = |t^\pm|Exp(i\zeta^\pm). \qquad (4)$$

$\zeta^\pm$ is the total phase accumulated by the normal mode as it propagates through the structure. The respective (power) transmittances are given by:

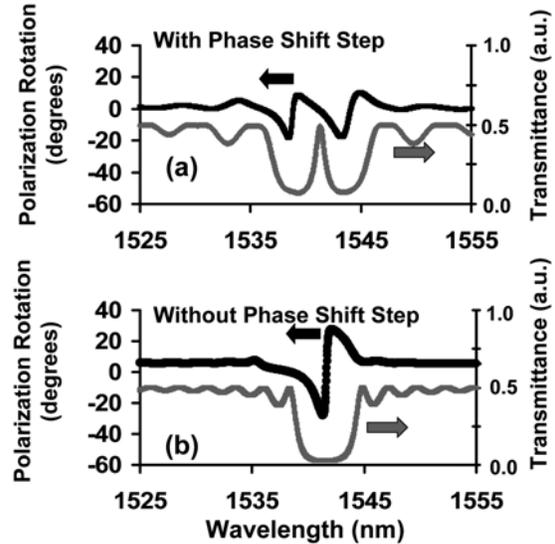

**Fig. 5** Calculated transmittance and polarization rotation for backscattering to the first-order waveguide mode for gratings with and without phase shift steps. The calculation is done by transfer-matrix simulation of the magnetophotonic structure under study.

$$T^{\pm} = |t^{\pm}|^2 \tag{5}$$

For a birefringence-free material, $\phi = 0$, $|E_y^{(n)}/E_x^{(n)}| = 1$, and the above relations reduce to the well-known expression for the Faraday rotation angle:

$$\theta = -\frac{1}{2}\Delta\zeta \tag{6}$$

For elliptical normal modes, the relation between the polarization rotation angle $\theta$ and the phase angle difference $\Delta\zeta$ is given by Eqs. (2) and (3). Large phase variations around the stop band result from the photonic bandgap structure. These are plotted in Fig. 4. The figure displays $\Delta\zeta$ obtained from the measured transmittance and polarization rotation in Fig. 3(b) using the formulation presented in this section. We also plot the result of a theoretical calculation of the phase for the same magnetophotonic crystal based on a transfer matrix formulation that treats the bandgap structure as a quarter-waveplate stack. Good agreement is seen with the experimental data.[7, 8, 15] The transfer matrix formulation is discussed in section 3.2, below.

It should be noted that one can reconstruct the polarization rotation angle $\theta$ from the normal-mode transmittance-phases $\zeta^+$ and $\zeta^-$.[15] Figure 5 shows the calculated rotation angle for the first-order mode backscattering under examination. Very good agreement is obtained with the measured polarization rotation in Fig. 3 for both cases, with and without phase shift step. However, it is clear that the relation between the normal-mode phase difference $\Delta\zeta$ and the resulting polarization rotation is no longer a linear relation as was the case for a birefringence-free system (Eq. (6)). An analysis of the kind presented in this section is required.

### 3.2 Elliptical normal-mode phases and group velocities

In examining the accumulated phases and the speed of propagation of the optical waves in the magnetophotonic crystals under study it is important to consider the normal modes separately. Each normal mode has a well-defined albeit different refractive index and each propagates at a different speed in the waveguide. In section 3.1 we analyzed the relation between normal mode phase differences and polarization rotation. We also verified that the analysis reproduces the measured polarization rotation in the stop bands. In this section we elaborate on the transfer matrix model and proceed to evaluate the spectral dependence of the speed and phase for each of the normal modes using this model.

In the transfer-matrix formulation the crystal is treated as a layered structure composed of quarter-wave thick plates.[7,8] The dielectric-tensor diagonal elements control the waveguide mode birefringence. Index contrast is modeled through the relative strength of the diagonal elements in adjacent layers, but keeping the overall birefringence uniform throughout the crystal along the propagation direction. This is a good approximation for the waveguides under study.

The transfer matrix method relates the amplitudes for forward ($A_x^f$, $A_y^f$) and backwards ($A_x^b$, $A_y^b$) propagating waves at the front ($z = 0$) and rear ($z = L$) of the stack by a matrix relation $A(L) = M\,A(0)$. Here $A = (A_x^f, A_y^f, A_x^b, A_y^b)$ and $M$ is a 4 x 4 matrix given by the product $P_n\,T_{n,\,n-1}\,P_{n-1}\,\ldots\,T_{2,\,1}\,P_1$, where $P_i$ propagates the amplitude across medium i, and $T_{i,\,i-1}$ relates the amplitudes across the interfaces between consecutive layers by imposing continuity of the fields from one region to the next.[7,8] Each quarter-wave plate has a thickness $h$,

$$h = \frac{\lambda_0}{4\left(\frac{\sqrt{\varepsilon_{xx}}+\sqrt{\varepsilon_{yy}}}{2}\right)}, \tag{7}$$

where $\lambda_0$ is the stop band center-wavelength, and $\varepsilon_{xx}$ and $\varepsilon_{yy}$ are the diagonal components of the dielectric tensor. The following dielectric tensors are used herein for the alternating layers, denoted A and B, to simulate the magnetophotonic crystal shown in Fig. 1:

$$\begin{pmatrix} \varepsilon_{xx}^A = 5.285 & \varepsilon_{xy}^A = 0.0027\,i & 0 \\ \varepsilon_{yx}^A = -0.0027\,i & \varepsilon_{yy}^A = 5.268 & 0 \\ 0 & 0 & \varepsilon_{zz}^A \end{pmatrix} \tag{8}$$

$$\begin{pmatrix} \varepsilon_{xx}^B = 5.242 & \varepsilon_{xy}^B = 0.0027\,i & 0 \\ \varepsilon_{yx}^B = -0.0027\,i & \varepsilon_{yy}^B = 5.225 & 0 \\ 0 & 0 & \varepsilon_{zz}^B \end{pmatrix} \tag{9}$$

Once the equivalent layered structure has been determined this way, the normal mode characteristics, including normal mode output phases $\zeta^\pm$ are extracted from the model. The polarization rotations in Fig. 5, computed on the basis of this transfer-matrix approach compare well with the experimental results and lend validity to the phases $\zeta^\pm$ extracted from these calculations.

Bendickson et. al. have proposed a formulation for the group velocity in a photonic crystal by defining an effective wave number $k$ equal to $\zeta/d$, where $\zeta$ is the total accumulated phase as the light propagates through the structure and $d$ is the physical thickness of

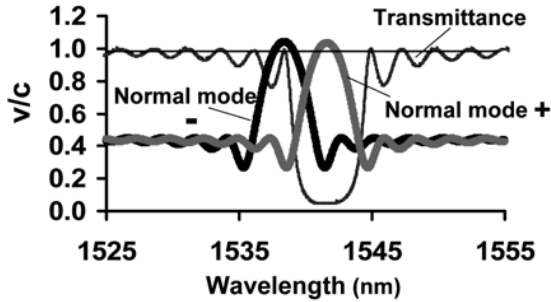

**Fig. 6** Calculated group velocities for the normal modes in a magnetophotonic crystal ridge waveguide without resonant cavity. Values are normalized to c, the speed of light in vacuum.

the same.[17] The group velocity of a propagating wave is then defined using the standard form

$$v_g = \frac{d\omega}{dk} \quad (10)$$

In terms of the accumulated phase $\zeta$ the group velocity is given by

$$v_g^\pm = \frac{d}{d\zeta^\pm/d\omega} \quad (11)$$

Here we adopt the same definition of group velocity as in [17]. Preliminary work in our research group indicates that this is a reasonable definition for a quarter-wave-plate stack photonic crystal, as long as the birefringence is uniform across the whole stack. We intend to elaborate further on the physical interpretation of the group velocity as defined here in a future publication.

Group velocities are calculated using the transfer-matrix model. Normal modes as defined by Eq. 1 are launched into the structure and their phases at the output-end computed numerically. Numerical differentiation then yields the group velocities. We point out that this transfer-matrix analysis also confirms that the normal modes concerned emerge in the same polarization state as the input except for an overall phase factor, as they should.

Figure 6 displays the normal-mode group velocities of the photonic crystal without phase-shift step. The modes resonate at different wavelengths because they have different effective indices. Away from resonance the group velocities show a rippled but otherwise flat response. Their value corresponds to $c/\bar{n}$, where $\bar{n}$ is the average refractive index of the material. At the edges of each resonance the photons slow down to $0.6 c/\bar{n}$. Notice that inside the stop band the photons speed up significantly, reaching superluminal speeds according to this calculation. A horizontal line at $v/c = 1$ is drawn to show this.

It is expected that for the magnetophotonic crystal depicted in Fig. 1 the phase-shift step will serve as a resonant cavity leading to the formation of standing waves and an additional time delay. This kind of photon trapping has often been cited as the origin of the enhanced Faraday rotation in magnetophotonic crystals. The question is: how much of the enhancement is due to standing-wave photon trapping and how much to actual changes in group velocity?

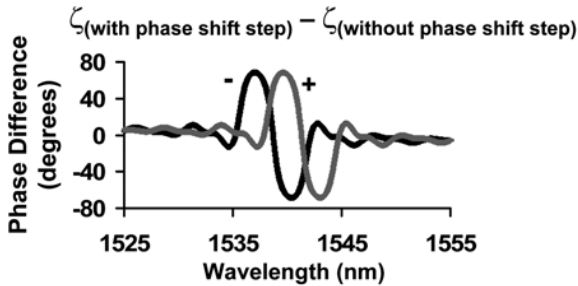

**Fig. 7** Normal mode difference in accumulated phase between propagation in the presence of a phase shift step and without a phase shift step. The difference results from group velocity differences in the spectral dependence and time delay engendered by photon tapping in the phase-shift-step resonant cavity.

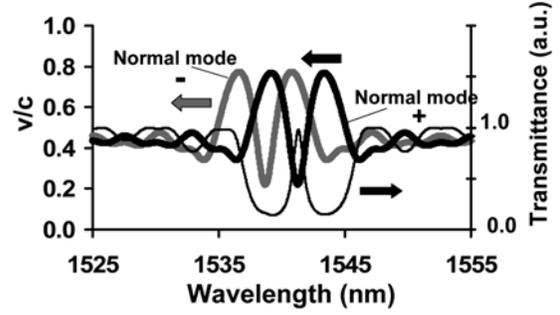

**Fig. 8** Calculated group velocities for the normal modes in a magnetophotonic crystal ridge waveguide with resonant cavity. Values are normalized to c, the speed of light in vacuum.

In order to begin addressing this interesting question, we have compared the accumulated phases of the normal modes in (a) the magnetophotonic crystal with phase-shift step and (b) without phase-shift step. The phase differences are plotted in Fig. 7. Our expectation is that if there is a significant time delay due to photon trapping there should be a significant phase difference near the resonance wavelength given by the Bragg condition. What we find is that although the phase difference is not zero at the resonance center-wavelengths, it is nevertheless small. The center-wavelengths are 1541.45 nm (+) and 1538.83 nm (-). In both cases the phase difference is -15° out of a maximum 68°. We consider this as an indication that standing-wave time delay is not the dominant effect here. Further investigation into this question is underway.

Figure 8 shows the calculated group velocities for both normal modes for the magnetophotonic crystal with resonant cavity phase-shift step. A significant slow-down is noted at both center-wavelengths, with speeds $v/c = 0.22$ or $v = 0.5 c/\bar{n}$.

## 4. Conclusions

We present a study of the polarization response and band edge effects in waveguide magnetophotonic crystals fabricated on $(Bi,Lu)_{(2.8 \pm 0.1)}Fe_{(4.6 \pm 0.2)}O_{(12.4 \pm 0.3)}$ films on (111) gadolinium gallium garnet (GGG) substrates. Fundamental to first-order mode coupling in photonic crystals with and without resonant cavity steps is analyzed and compared. A formulation is presented that relates the polarization response to normal mode phase differences in these linearly-birefringent polarization rotators. In contrast to the Faraday Effect in non-birefringent systems a non-linear relation exists between polarization rotations and normal mode phase differences. Transfer-matrix modeling is used to extract the normal mode phase retardation in these magnetophotonic crystals. This analysis yields evidence of optical slow-down at the Bragg resonance center-wavelength for resonant cavity steps to one-half the speed of light in the original dielectric medium. In contrast, significant speed-up is found in the Bragg grating without phase-shift step.

**Acknowledgments** This work was supported by the National Science Foundation under grant ECS-0520814 and the Department of Defense under grant DAAD 17-03-C-0115. The research reported in this document was performed in connection with contract DAAD17-03-C-0115 with the U.S. Army Research Laboratory through the Defense Advanced Research Projects Agency. The views and conclusions contained in this document are those of the authors and should not be interpreted as presenting official policies or position, either expressed or implied, of the U.S. Army Research Laboratory, DARPA, or the U.S. Government unless so designated by other authorized documents.

Citation of manufacturer's or trade names does not constitute an official endorsement or approval of the use thereof. The U.S. Government is authorized to reproduce and distribute reprints for Government purposes notwithstanding any copyright notation hereon.